\documentclass[useAMS,usenatbib]{mn2e}
\usepackage{aas_macros,graphicx,times,multirow,amsmath}
\usepackage[]{color}

\title[Neutron star spin evolution in the Be/X-ray binary SXP 1062]{Probing the neutron star spin evolution in the
young SMC Be/X-ray binary SXP 1062}
\author[S.B. Popov \& R. Turolla]{S.B.~Popov$^{1,2}$\thanks{E-mail: sergepolar@gmail.com}
and R.~Turolla$^{2,3}$
\smallskip\\
$^1$Sternberg Astronomical Institute, Moscow State University, Moscow, Russia 119992\\
$^2$Dipartimento di Fisica e Astronomia, Universit\`{a} di Padova, via Marzolo 8, I-35131 Padova, Italy\\
$^3$Mullard Space Science Laboratory, University College London, Holmbury St. Mary, Dorking, Surrey RH5 6NT, UK\\
}
\date{Accepted \ldots. Received \ldots; in
original form \ldots} \pagerange{\pageref{firstpage}--\pageref{lastpage}} \pubyear{2011}

\def\LaTeX{L\kern-.36em\raise.3ex\hbox{a}\kern-.15em
    T\kern-.1667em\lower.7ex\hbox{E}\kern-.125emX}

\def\xmm {\emph{XMM-Newton}}

\def\po {P_{orb}}
\def\src {SXP 1062}

\def\lum {\mbox{erg s$^{-1}$}}

\begin{document}

\label{firstpage}
\maketitle
\begin{abstract}
The newly discovered Be/X-ray binary in the Small Magellanic
Cloud, \src, provides the first example of a robust association
with a supernova remnant (SNR). The short age estimated for the
SNR qualifies \src\ as the youngest known source in its class,
$\tau\approx 10^4\ \textrm{yr}$. As such, it allows to test
current models of magneto-rotational evolution of neutron stars in
a still unexplored regime. Here we discuss possible evolutionary
scenarios for \src\ in the attempt to reconcile its long spin
period, $P=1062\ \textrm{s}$, and short age. Although several
options can be considered, like an anomalously long initial period
or the presence of a fossil disc, our results indicate that \src\
may host a neutron star born with a  large initial magnetic field,
typically in excess of $\sim 10^{14}\, {\rm G}$, which then
decayed to $\sim 10^{13}\, {\rm G}$.

\end{abstract}

\section{Introduction}
\label{intro}

Be/X-ray binaries (or BeXBs for short) form a subclass of the
high-mass X-ray binaries (HMXBs) in which the neutron star (NS)
companion is a Be star, a spectral class B giant/subgiant with
emission lines and large IR flux. These peculiar properties are
explained in terms of an equatorial disc, formed by matter lost by
the rapidly rotating Be star. X-ray emission is believed to be
powered by accretion of material in the equatorial disc onto the
NS \cite[see e.g.][for a recent review]{reig11}.

BeXBs are both transient and persistent X-ray sources. Transient
systems are characterized by type I-II outbursts during which
their flux increases by a factor 10--$10^4$ over the quiescent
level. They typically contain a not-too-slow NS ($P\la 100\, {\rm
s}$) on a moderately eccentric orbit, $\po\la 100\, {\rm d}$,
$e\ga 0.2$. On the other hand, persistent BeXBs exhibit a rather
flat lightcurve, lower X-ray luminosity ($L\approx
10^{34}$--$10^{35}\, \lum$), longer spin and orbital periods
\cite[$P\ga 200\, {\rm s}$, $\po\ga 200\, {\rm d}$; see
again][]{reig11}.

There are presently about 30 well-established BeXBs in the Galaxy,
plus $\sim 40$ candidates. In addition, $\sim 50$ sources (plus
$\sim 20$ candidates) are known in the Small Magellanic Cloud
(SMC)\footnote{Figures from \cite{reig11} and the BeXB online
catalogue at http://xray.sai.msu.ru/\textasciitilde
raguzova/BeXcat; see also \cite{ragpo05}.}. Very recently
\cite{henbru11} reported the discovery of a new BeXB in the Wing
of the SMC. The new source (\src) has the typical properties of a
persistent BeXB: a B0-0.5(III)e+ donor, $L\sim 6\times 10^{35}\,
\lum$, $P\sim 1062\, {\rm s}$, and an orbital period $\po\sim
300\, {\rm d}$, as estimated from the Corbet diagram
\cite[][]{corb09}. What makes \src\ to stand out amongst its
kinship is its (likely) association with a supernova remnant
(SNR). BeXB-SNR associations have been already reported (again in
the SMC) but in all previous cases they appear uncertain
\cite[][]{hugsm94,coe00}. In the case of \src\ the association
looks robust and allows for the first time to estimate the NS age
in a BeXB from that of the parent SNR, $\tau\sim 2$--$4\times
10^4\, {\rm yr}$ \cite[][]{henbru11}. The suggested association of
\src\ with a SNR has been further strengthened by a reanalysis of
the same \xmm\ datasets, supplemented with optical and radio
observations, by \cite{hab11}. Their estimate for the SNR age,
$1.6\times 10^4$ yr, is even shorter than, albeit fully compatible
with, that of \cite{henbru11}. \cite{hab11} were also able to
measure the source period evolution, which results in a positive
(i.e. spin-down) rate $\dot P\sim 3\times 10^{-6}\, {\rm
s\,s}^{-1}\sim 95\, {\rm s\,yr}^{-1}$.

The long spin periods ($P \ga 1000\, {\rm s}$) of some persistent
BeXBs have been for a long time a major issue. According to the
standard picture, there are four stages in the spin evolution of a
neutron star embedded in a medium: ejector, propeller, accretor
and georotator\footnote{The last one is of no concern here, since
it occurs under specific conditions, hardly met in HMXBs.}
\cite[e.g.][]{lip92}. Once the NS entered the accretor stage after
a short propeller phase, its spin period quickly settles at an
equilibrium value, $P_{eq}$. In the conventional model, based on
\cite{davpri81} results, the star dipole field should be $B\ga
10^{14}\, {\rm G}$ to have $P_{eq}>10^3 \, {\rm s}$, unless the
accretion rate is orders of magnitude below what needed to account
for the observed X-ray luminosity. Since observations support the
presence of a NS with standard magnetic field (at least in some
BeXBs), the subsonic propeller stage, which can delay the onset of
accretion until a much longer period is reached \cite[][]{iks07},
has been invoked to explain ultra-slow BeXBs \cite[see][]{reig11}.
More recent investigations of wind-fed accretion onto magnetized
NSs indicate, however, that the equilibrium period can be as high
as $\sim 1000\, {\rm s}$ even for $B\approx 10^{12}$--$10^{13}\,
{\rm G}$ and  $\dot M\approx 10^{16}\, {\rm g\,s}^{-1}$, as
expected in BeXBs \cite[][]{shak11}.

Whatever the details of the braking torques, the previous argument
implicitly relies on the assumption that the present age of the
source is long enough for the NS to have entered the propeller
stage. Unless the accretion rate is way above typical, a NS in a
BeXB with $B\ga 10^{12}\, {\rm G}$ starts its evolution in the
ejector (or pulsar) phase. Its duration can be roughly estimated
as $\tau_{ej}\ga 10^6 (B/10^{12}\, {\rm G})^{-1}(\dot M/10^{15}\,
{\rm g\,s}^{-1})^{-1/2}\, {\rm yr}$ (see Section~\ref{spindown}).
This is comfortably below (by a factor $\approx 10$) the lifetime
of the Be companion, so there is ample room for the binary to
start an accretion-powered X-ray stage. In the case of \src,
however, it would be impossible for the NS to enter the propeller
stage (and hence to become an accretor) in a time as short as a
few $\times  10^4\, {\rm yr}$, the estimated SNR age, for typical
values of $B$ and $\dot M$. The accretion rate in \src\ is $\dot
M=L/\eta c^2\sim 6\times 10^{15} \, {\rm g\,s}^{-1}$ for an
efficiency $\eta=0.1$, so this points to a highly magnetized NS,
with an initial magnetic field substantially above $10^{12}\, {\rm
G}$.

\section{Spin evolution in \src}
\label{spindown}

\subsection{Spin-down torques}
\label{spindowntorq}

In the standard picture of NS spin evolution \cite[see
e.g.][]{lip92}, the transitions among the different stages are
regulated by the relative values of some characteristic radii. In
the ejector phase the light cylinder radius, $R_l=cP/2\pi\sim 5
\times 10^9 \, P\, {\rm cm}$, is typically smaller than the
gravitational capture radius, $R_G=2GM/V^2\sim 4\times 10^{11} \,
V_{300}^{-2}\, {\rm cm}$, where $V$ is the velocity (and $\rho$
the density) of matter far from the star\footnote{Hereafter
$V_{300}\equiv (V/300\, {\rm km\,s}^{-1})$, $P_{orb\, 300}\equiv
(P_{orb}/300\, {\rm d})$, $P_{1000}\equiv (P/1000\, {\rm s})$ and
$N_X$ is used for the quantity $N$ in units of $10^X\, {\rm
cgs}$.}. A NS with $M_{NS}=1.4\, M_\odot$, $R_{NS}=10\, {\rm km}$
and moment of inertia $I=10^{45}\, {\rm g\,cm^2}$ is assumed
henceforth. The transition to the propeller stage occurs when the
ram pressure, $P_{dyn}=\rho V^2/2$, balances the outgoing flux of
electromagnetic waves and relativistic particles, $P_{PSR}=\dot
E/(4\pi R^2c)$, at $R_G$ ($\dot E$ is the rotational energy loss
rate of the pulsar). Once matter crosses $R_G$ it is
gravitationally captured and quickly reaches the light cylinder
radius switching off pulsar emission, since the dynamical pressure
(assuming a nearly spherical flow) rises as $R^{-5/2}$, while the
relativistic momentum flux goes like $R^{-2}$.

The critical period for the transition follows by requiring that
$P_{dyn}(R_G)=P_{PSR}(R_G)$ together with the
standard expression for magneto-dipole losses,
$\dot E=8\pi^4B^2R_{NS}^6\sin^2\alpha/(3c^3P^4)$ and mass
conservation, $4\pi R_G^2\rho V=\dot M$,

\begin{equation}
\label{pejec}
P_{ej}=2 \pi \left( \frac43 \frac{B^2R_{NS}^6}{\dot M V
c^4}\right)^{1/4} \sim 0.31 \, V_{300}^{-1/4} \dot M_{16}^{-1/4}
B_{12}^{1/2} \, {\rm s};
\end{equation}
since the NS is not accreting, here $\dot M$ is more properly defined as the matter flow rate in the
surrounding medium and we assumed a nearly orthogonal rotator ($\sin\alpha\sim 1$).
From the magneto-dipole formula, assuming constant $B$ and very short initial period, the time it takes the NS
to spin down to $P_{ej}$, i.e. the duration of the ejector stage, can be evaluated

\begin{equation}
\label{ejectime}
\tau_{ej} =\frac{3Ic^3 P_{ej}^2}{16\pi^2 B^2
R_{NS}^6}\sim
1.5\, \dot M_{16}^{-1/2}  V_{300}^{-1/2} B_{12}^{-1}\, {\rm Myr}.
\end{equation}

The dipole field in a wind-fed NS has been estimated by
\cite{shak11} under the assumption that the star is spinning at
the equilibrium period \cite[see also][]{post11},

\begin{equation}
\label{bshak} B_{12} \sim 8.1\, \dot M_{16}^{1/3}
V_{300}^{-11/3}\left(\frac{P_{1000}}{P_{orb\,
300}}\right)^{11/12}\, .
\end{equation}
The previous expressions give $\tau_{ej}\sim 0.2\, {\rm Myr}$ for
\src, about an order of magnitude larger than the estimated SNR
age. Our reference value, $V\sim 300\, {\rm km\,s}^{-1}$, is a
conservative estimate for the typical velocity of matter ejected
from hot stars in binaries \cite[see][]{raglip98}. Actually, since
Eq. (\ref{bshak}) depends rather strongly on $V$, higher
velocities would result in a lower magnetic field and longer
$\tau_{ej}$.

Within this framework, an obvious possibility to shorten the
ejector phase in \src\ is to invoke a higher dipole field.
However, if the present field is that given by Eq. (\ref{bshak})
this implies that $B$ must have been stronger in the past and then
decayed to its present value. The issue of magnetic field
evolution in isolated NSs and its observable consequences has been
recently addressed in detail, especially in connection with
strongly magnetized objects, or magnetars \cite[e.g.][and
references therein]{pons09, popov10, turolla11}. In these studies
it is assumed that magnetic field decay takes place in the stellar
crust and is driven by Hall/Ohmic diffusion. Both processes are
strongly density- and temperature- dependent, so magnetic and
thermal evolution are necessary coupled, and  a multidimensional
numerical approach is needed \cite[][]{pons09}.

Since we are interested in tracing the spin evolution of \src\
prior it entered the accretor stage, we can treat the NS as
isolated for the sake of magnetic evolution. In order to avoid
complex numerical simulations, we adopt the simplified approach of
\cite{aguil08} in which the dipole magnetic field decay is
described by the analytical law
\begin{equation}
\label{bdecay}
B(t)=\frac{B_0\exp{(-t/\tau_O)}}{1+(\tau_O/\tau_H)[1-\exp{(-t/\tau_O)}]}+B_{fin},
\end{equation}
where $B_0$ is the initial field, $B_{fin}$ is the relic field,
$\tau_O$ and $\tau_H$ are the characteristic timescales of Ohmic
and Hall decay, respectively. The period evolution in the ejector
stage is given by the (generalized) magneto-dipole formula

\begin{equation}
\label{ejtorque}
\dot P=\frac{8\pi^2B^2R_{NS}^6(1+\sin^2\alpha)}{3Ic^3P}
\end{equation}
\cite[][]{spit06}, where $B$ is given by Eq. (\ref{bdecay}).

The NS enters the propeller phase as soon as the dynamical
pressure exerted by the incoming material overwhelms the pulsar
momentum flux at the gravitational radius, as discussed above. If
no stable equilibrium exists, matter will reach the light cylinder
radius on a free-fall timescale, and proceed inwards if the ram
pressure overcomes the magnetic pressure of the dipole field,
$P_{mag}=B^2/8\pi$. \cite{eksialpar05} have shown that stable
matter configurations can be present even if the inner boundary of
the flow, $R_{in}$, is outside $R_l$. This occurs when
$R_{in}<R_{crit}=F(\alpha) R_A$, with $F(\alpha)\geq 1$ a function
of the magnetic tilt angle and\footnote{Although $R_A$ formally
coincides with the  Alfv\'en radius, we remark that the latter
exists only inside the light cylinder radius. Nonetheless, the
present definition is correct, provided that $R_A$ is not
associated to the equilibrium of dynamical and magnetic pressure.}
$R_A=(1/2)B^{4/7}R_{NS}^{12/7}/(8GM_{NS}\dot M^2)^{1/7}$. For
$R_l<R_{in}<R_{crit}$, the pulsar is still active and the star is
spun down by magneto-dipole torque. Under these conditions
\cite{eksialpar05} provide an expression for $R_{in}$ in terms of
$\alpha$ and $R_A$. When $R_{in}<R_l$, instead, the flow is
truncated at the Alfv\'en radius, where $P_{dyn}=P_{mag}$, the
pulsar is turned off and only propeller torques act.

The propeller physics is very complicated and no unanimous
consensus exists about the form of the torque. However, since
spin-down is expected to be very efficient in the propeller phase,
its duration is quite short. The period rapidly increases and the
corotation radius, $R_{co}=[GM_{NS}/(4\pi^2)]^{1/3}P^{2/3}$, moves
quickly outwards until it matches the  Alfv\'en radius, at which
point accretion begins and the period freezes at the equilibrium
value. While the latter depends on the propeller/accretion
mechanisms, and several options have been discussed (see Sec.
\ref{intro}), the duration of the ejector phase does not. For this
reason we consider here only the ``maximally efficient torque''
\cite[see e.g.][]{franci02}, as an illustrative example:

\begin{equation}
\label{maxtorque}
\dot P = -\dot M R_{in}^2\left[\Omega_K(R_{in})-2\pi/P\right]\frac{P^2}{\pi I}
\end{equation}
where $\Omega_K$ is the keplerian angular velocity.

Finally, to follow the period evolution in the accretor stage,
$R_{in}< R_{co}$, we assume the settling accretion regime recently
proposed by \citet[][see also \citealt{post11}]{shak11}

\begin{equation}
\label{shaktorque}
\dot P = -\left[A\dot M_{16}^{(3+2n)/11} - C\dot M_{16}^{3/11} \right]\frac{P^2}{2\pi I}.
\end{equation}
Here $A\sim 2.2\times 10^{32}K_1(B_{12}R_{NS\,
6})^{1/11}V_{300}^{-4}P_{orb\, 300}^{-1}$, $C\sim 5.4\times
10^{31} K_1(B_{12}R_{NS\, 6})^{13/11}P_{1000}^{-1}$ (both in cgs
units); the constant $K_1$ and the index $n$ were fixed to 40 and
2, respectively. The accretion torque can produce either spin-up
or spin-down and it is interesting to note that in the spin-down
phase $P$ increases exponentially, with a typical timescale
$\approx 100B_{12}^{-13/11}\dot M_{16}^{-3/11}\, {\rm yr}$.

\subsection{Results}
\label{res}

\begin{figure} \resizebox{\hsize}{!}{\includegraphics[angle=0]{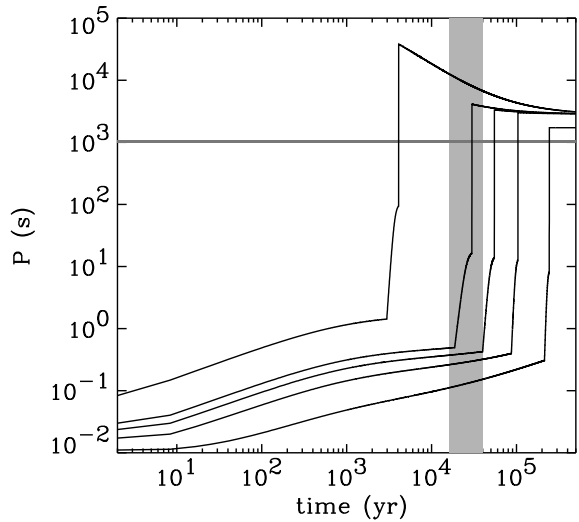}}
\caption{\label{perplot} The spin period evolution for
$\alpha=10^\circ$, $\tau_H=10^3\, {\rm yr}$ and $B_0=4\times
10^{14}$, $10^{14}$, $7\times 10^{13}$, $4\times 10^{13}$,
$10^{13}\, {\rm G}$ (solid lines, from top to bottom). The shaded
areas mark the age (vertical strip) and period (horizontal strip)
of \src\ with the respective uncertainties. }
\end{figure}

We solved numerically the equation for the period evolution in the
three stages (Eqs. [\ref{ejtorque}], [\ref{maxtorque}],
[\ref{shaktorque}]) starting from $t_0=0.01\, {\rm yr}$ with an
initial period $P_0=0.01 \, {\rm s}$. The accretion rate was fixed
to the value derived from current observations, $\dot M =6\times
10^{15}\, {\rm g/s}$, together with $\po=300\, {\rm d}$, $V=300\,
{\rm km\,s}^{-1}$, $\tau_O=10^6\, {\rm yr}$ and $B_{fin}=8\times
10^{12}\, {\rm G}$. Several cases were then computed varying  the
magnetic dipole angle $\alpha$, the Hall timescale $\tau_H$ and
especially the initial field $B_0$. Figure \ref{perplot}
illustrates the results for $\alpha=10^\circ$, $\tau_H=10^3\, {\rm
yr}$ and $B_0=4\times 10^{14}$, $10^{14}$, $7\times 10^{13}$,
$4\times 10^{13}$, $10^{13}\, {\rm G}$. A common feature to all
evolutionary tracks is a very rapid propeller stage which is
followed by an even more rapid spin-down phase as the NS enters
the accretion regime. The period then saturates at its equilibrium
value. The decrease in $P$ seen at later times for the larger
fields is due to the dependence of $P_{eq}$ on $B$ and to the fact
that the field is still decaying (see \ Sec. \ref{spindowntorq}).
The time variation of $B$ is also responsible for the deviations
of $P(t)$ from a power-law in the later ejector phase, which are,
again, more prominent for large $B$.

The main information Figure \ref{perplot} conveys is that a quite
large initial field is required in order for \src\ to enter the
propeller phase (and quickly start accreting) in a time as short
as a few $\times 10^4 \, {\rm yr}$. For the case shown here it has
to be $B_0\ga 10^{14}\, {\rm G}$ for this to occur. This result is
not very sensitive to the actual choice of $\alpha$ and $\tau_H$.
Both increasing $\alpha$ and decreasing $\tau_H$ results in a
somehow higher value of the  minimum initial field, which is
however in all cases in the range $\sim 10^{14}\,$--$\,4\times
10^{14}\, {\rm G}$. The conclusion that \src\ harbours an
initially strongly magnetized NS seems therefore quite robust.

We stress that our main goal is not to reproduce within the
current model the observed value of $P$ at the present age.
Although, for completeness,  we followed the spin evolution also
in the propeller and accretor stages, the treatment we employed is
necessary approximated and contains some degrees of arbitrariness
in the choice of the propeller/accretor torques. Our computed
evolution past the ejector phase has mainly an illustrative
purpose and has to be taken with caution.

\section{Discussion}

The quite short age implied by the association of the newly
discovered BeXB in the SMC \src\ with a SNR \cite[][]{henbru11,
hab11} raises a number of questions about the properties of the
neutron star and its evolutionary status. Normally, one expects
that accreting X-ray pulsars spin close to their equilibrium
period. However, for a such a young system this appears far from
granted. \cite{hab11} reported a large spin-down rate for \src,
which may suggest that $P_{eq}$ has not been reached yet.
According to the standard evolutionary scenario \cite[][]{lip92},
the (maximum) spin-down rate in the accretor stage is $\dot P\sim
2\pi B^2R_{NS}^6/(GMI)$ which implies $B\approx 3\times 10^{14}\,
{\rm G}$ for $\dot P\approx 100\, {\rm s\,yr}^{-1}$. On the other
hand, if Eq. (\ref{shaktorque}) is used to estimate the magnetic
field, a much lower value is obtained, $B\approx 10^{13}\, {\rm
G}$, very close to what is predicted assuming that the source
spins at the equilibrium period (see Eq. [\ref{bshak}]). This
supports a picture in which the NS actually rotates close to
$P_{eq}$. A further argument in favor of this is the very short
duration of the spin-down phase in the accretor stage, $P/\dot
P\la 100\, {\rm yr}$, which makes it very unlikely to catch the
source in this state. Our conclusion is that both the young age
and the large spin-down rate of \src\ argue in favor of an
initially highly magnetic NS which experienced field decay.

Details of evolution past the ejector stage are uncertain, so a
fine tuning of the model discussed in Sec. \ref{spindown} is
entirely premature. Still, some general considerations can be
made. The main one is that the timescale required for reaching
$P_{eq}$ after the NS left the ejector stage is very short. A
further point to be stressed is that the zero, or negative, period
derivative at $P=P_{eq}$ (see Figure \ref{perplot}) does not
account for variations in $\dot M$ due to orbital motion in the
BeXB and irregularities in the wind. This can result in quite
large values of $\dot P$ and rapid changes in the period
derivative which are indeed measured in wind-fed X-ray binaries
\cite[see e.g.][]{bild97}. In this respect the spin-down rate in
\src\ is large but not exceptionally so. For example, GX~301-2 is
known to have a larger value of $\dot P$ \cite[][]{koh97}.
\cite{hab11} measured the average period derivative over a
time-span less than a month, $\sim$ one tenth of the orbital
period. Fluctuations in $\dot P$ can occur on a timescale of days,
so a more complete dataset is definitely needed to assess the real
nature of the period variation.

Alternative scenarios to explain the long period and short age of
\src\ can be envisaged. For instance, \cite{hab11} suggested that
the NS could have been born with an initial period much longer
than $\sim 0.01\, {\rm s}$. The value of $P_0$ can be evaluated by
requiring that the transition period given by Eq. (\ref{pejec}) is
reached in less than the source age, and it turns out to be $\sim
1\, {\rm s}$ for the $B$-field corresponding to $P_{eq}$ [Eq.
(\ref{bshak})]. If this is the case no field decay is required.
Although possible, no compelling observational evidence for such
long initial periods in NSs exists. There are hints that some
central compact objects (CCOs) in SNRs may have a current period
very close to the initial one, $P_0\approx 0.1\, {\rm s}$. These
sources, however, are suspected to host weakly magnetized NSs
\cite[$B\approx 10^{11}\, {\rm G}$; e.g.][]{gotthalp10}. Whether
the low field and the long period are related, depending on the
conditions under which the NS formed, is still unclear. Still, the
estimated value of $B$ in \src\ is much higher and the initial
period required in the present case is more than one order of
magnitude longer. Very recently \cite{knig11} presented evidence
that the population of BeBXs is bimodal, with the two
sub-populations having distinct typical spin and orbital periods,
and eccentricities. They suggest that the NSs in the two groups
are produced through different channels, iron-core-collapse and
electron capture supernovae, with the long spin period sources
associated to the former channel. Whether this may result in a
population of long $P_0$ NSs is a still open question.

Another possibility is that the NS in \src\ could, at least in the
early stages following its formation, have been surrounded by a
debris (or fossil) disc, left by the parent supernova explosion.
The existence of such a disc could also lead to rapid spin-down
and large period. This, in addition to a super-strong field, was
suggested to explain the ultra-long period ($\sim 6.67\, {\rm
hr}$) in the enigmatic source RCW~103 \cite[][]{deluca06,li07}.
Although this remains a possibility worth of further
investigations, preliminary calculations, based on the model by
\citet[][see also \citealt{esp11}]{li07}, indicate that a large
initial field ($B\ga 10^{14}\ {\rm G}$) is still required, unless
the disc is quite massive. For $M_{disc}(0)\sim 10^{-2}\ M_\odot$,
the NS can enter the propeller stage earlier than $\sim 10^4\ {\rm
yr}$ also for $B\sim 10^{13}\ {\rm G}$.

If \src\ indeed contains an initially strongly magnetized neutron
star, then studies of this system can shed light on the origin of
magnetars. In the standard scenario \cite[][]{dt92} super-strong
fields are generated via dynamo processes. This requires rapid
rotation of the proto neutron star. Primary components in
high-mass binary systems are not expected to form rapidly rotating
cores at the end of their lives \cite[][]{pp06, bp09}. Up to now
no strongly magnetized compact objects have been identified in
binary systems with certainty. \cite{chashkina11}  have recently
derived estimates of the $B$-field in HMXBs using \cite{shak11}
formula and no evidence of ultra-high fields was found. It has
been suggested that supergiant fast X-ray transients \cite[SFXTs,
see e.g.][for a review]{sid11} may host a magnetar
\cite[][]{boz08}. More recently, \cite{tor11} reported
magnetar-like behaviour from the peculiar binary LS I +61 303.
However, no definite confirmation has been given yet, and, in this
respect, \src\ may be a unique example. The existence of a
(evolved) magnetar in a high-mass binary system will pose new
challenges on the origin of such neutron stars.

\section*{Acknowledgments}
SBP gratefully acknowledges a Visiting Scientist grant from the
University of Padova and the Department of Physics and Astronomy
of the University of Padova for the hospitality during the period
this investigation was carried out. We thank P. Esposito, F.
Haberl and V. H\'{e}nault-Brunet for some helpful comments on the
manuscript, and an anonymous referee for a number of insightful
suggestions. RT is partially supported by INAF/ASI under contract
I/009/10/0. SBP is also supported by RFBR grant 10-02-00059.

\bsp

\label{lastpage}


\begin{thebibliography}{}

\bibitem[\protect\citeauthoryear{Aguilera et al.}{2008}]{aguil08}
Aguilera D.N., Pons J.A., Miralles J.A., 2008, A\&A, 486, 255

\bibitem[\protect\citeauthoryear{Bildstein et al.}{1997}]{bild97}
Bildsten L., et al., 1997. ApJS, 113, 367

\bibitem[\protect\citeauthoryear{Bogomazov \& Popov}{2009}]{bp09}
Bogomazov A.I., Popov S.B., 2009, Astron. Rev., 53, 325

\bibitem[\protect\citeauthoryear{Bozzo, Falanga \& Stella}{2008}]{boz08}
Bozzo E., Falanga M., Stella L., 2008, ApJ, 683, 1031

\bibitem[\protect\citeauthoryear{Coe, Haigh \& Reig}{2000}]{coe00}
Coe M.J, Haigh N.J., Reig P., 2000, MNRAS, 314, 290

\bibitem[\protect\citeauthoryear{Corbet et al.}{2009}]{corb09}
Corbet R.H.D., Coe M.J., McGowan K.E., Schurch M.P.E., Townsend
L.J., Galache J.L., Marshall F.E., 2009, in van Loon J., Oliveira
J.M., eds, Proc. IAU Symp. 256, The Magellanic System: Stars, Gas,
and Galaxies. Kluwer, Dordrecht, p. 361

\bibitem[\protect\citeauthoryear{Chashkina \& Popov}{2012}]{chashkina11}
Chashkina A., Popov S.B., 2012, New Astronomy, submitted
(arXiv1112.1123)

\bibitem[\protect\citeauthoryear{Davies \& Pringle}{1981}]{davpri81}
Davies R.E., Pringle J.E., 1981, MNRAS, 196, 209

\bibitem[\protect\citeauthoryear{De Luca et al.}{2006}]{deluca06}
De Luca A., Caraveo P.A., Mereghetti S., Tiengo A., Bignami G.F., 2006, Science, 313, 814

\bibitem[\protect\citeauthoryear{Duncan \& Thompson}{1992}]{dt92}
Duncan R.C., Thompson C., 1992, \apjl, 392, 9

\bibitem[\protect\citeauthoryear{Ek\c{s}i \& Alpar}{2005}]{eksialpar05}
Ek\c{s}i K.Y., Alpar M.A., 2005, ApJ, 620, 390

\bibitem[\protect\citeauthoryear{Esposito et al.}{2011}]{esp11}
Esposito P., Turolla R., De Luca A., Israel G.L., Possenti A.,
Burrows D.N. 2011, MNRAS,

\bibitem[\protect\citeauthoryear{Francischelli \& Wijers}{2002}]{franci02}
Francischelli G.J., Wijers R.A.M.J. 2002, preprint (astro-ph/0205212)

\bibitem[\protect\citeauthoryear{Gotthelf \& Halpern}{2010}]{gotthalp10}
Gotthelf E.V., Halpern J.P., 2010, ApJ, 709, 436

\bibitem[\protect\citeauthoryear{Haberl et al.}{2012}]{hab11}
Haberl F., Sturm R., Filipovi\'c M.D., Pietsch W., Crawford E.J.,
2012, A\&A, 537, L1

\bibitem[\protect\citeauthoryear{H\'{e}nault-Brunet et al.}{2012}]{henbru11}
H\'{e}nault-Brunet V. et al., 2012, MNRAS,
doi:10.1111/j.1745-3933.2011.01183.x (arXiv:1110.640)

\bibitem[\protect\citeauthoryear{Hughes \& Smith}{1994}]{hugsm94}
Hughes J.P., Smith R.C., 1994, AJ, 107, 1363

\bibitem[\protect\citeauthoryear{Ikhsanov}{2007}]{iks07}
Ikhsanov N.R., 2007, MNRAS, 375, 698

\bibitem[\protect\citeauthoryear{Knigge, Coe \& Podsiadlowski}{2012}]{knig11}
Knigge C., Coe M.J., Podsiadlowski P., 2012, Nature, in press
(arXiv:1111.205)

\bibitem[\protect\citeauthoryear{Koh et al.}{1997}]{koh97}
Koh D.T. et al., 1997, ApJ 479, 933


\bibitem[\protect\citeauthoryear{Li}{2007}]{li07}
Li X.-D., 2007, \apjl, 666, L81

\bibitem[\protect\citeauthoryear{Lipunov}{1992}]{lip92}
Lipunov V.M., 1992, Astrophysics of Neutron Stars, Berlin, Springer-Verlag

\bibitem[\protect\citeauthoryear{Pons, Miralles \& Geppert}{2009}]{pons09}
Pons J.A., Miralles J.A., Geppert U., 2009, A\&A, 496, 207

\bibitem[\protect\citeauthoryear{Popov \& Prokhorov}{2006}]{pp06}
Popov S.B., Prokhorov M.E., 2006, MNRAS, 367, 732

\bibitem[\protect\citeauthoryear{Popov et al.}{2010}]{popov10}
Popov S.B., Pons J.A., Miralles J.A., Boldin P.A., Posselt B., 2010, MNRAS, 401, 2675

\bibitem[\protect\citeauthoryear{Postnov et al.}{2011}]{post11}
Postnov K., Shakura N., Kochetkova A., Hjalmarsdotter L., 2011, in the proceedings of the workshop ``The Extreme
and Variable High X-ray Sky'', September 19-23, 2011, Chia Laguna, Sardegna, Italy (arXiv:1110.1382)

\bibitem[\protect\citeauthoryear{Raguzova \& Lipunov}{1998}]{raglip98}
Raguzova N.V., Lipunov V.N., 1998, A\&A, 340, 85

\bibitem[\protect\citeauthoryear{Raguzova \& Popov}{2005}]{ragpo05}
Raguzova N.V., Popov S.B., 2005, Astron. Astrophys. Transactions, 24, 151

\bibitem[\protect\citeauthoryear{Reig}{2011}]{reig11}
Reig P., 2011, Ap\&SS, 322, 1

\bibitem[\protect\citeauthoryear{Shakura et al.}{2012}]{shak11}
Shakura N., Postnov K., Kochetkova A., Hjalmarsdotter L., 2012,
MNRAS, in press (arXiv:1110.3701)

\bibitem[\protect\citeauthoryear{Sidoli}{2011}]{sid11}
Sidoli L ., 2011, in the proceedings of the workshop ``The Extreme
and Variable High X-ray Sky'', September 19-23, 2011, Chia Laguna, Sardegna, Italy (arXiv:1111.5747)

\bibitem[\protect\citeauthoryear{Spitkovsky}{2006}]{spit06}
Spitkovsky A., 2006, ApJ, 648, L51

\bibitem[\protect\citeauthoryear{Torres et al.}{2012}]{tor11}
Torres D.F., Rea N., Esposito P.,  Li J., Chen Y., Zhang S., 2012,
ApJ, 744, 106

\bibitem[\protect\citeauthoryear{Turolla et al.}{2011}]{turolla11}
Turolla R., Zane S., Pons J.A., Esposito P., Rea N., 2011, ApJ, 740, 105

\end{thebibliography}
\end{document}